\documentclass[11pt]{article}
\usepackage{moriond,epsfig}

\def\be{\begin{equation}}
\def\ee{\end{equation}}
\def\bea{\begin{eqnarray}}
\def\eea{\end{eqnarray}}

\def\rtsp{\ensuremath{\sqrt{s^{\prime}}}}
\def\deb{\ensuremath{\Delta E_{\rm beam}}}
\begin{document}
\vspace*{4cm}

\title{DETERMINATION OF THE LEP BEAM ENERGY USING Z RECOIL EVENTS}

\author{ CHRISTIAN ROSENBLECK }

\address{
   III. Physikalisches Institut A,
   RWTH Aachen,
   D-52062 Aachen,
   Germany
}

\maketitle

\abstracts{
  The precise knowledge of the beam energy, $E_{\rm beam}$,  at the LEP collider is important to reduce
  the systematic uncertainty on the W mass.
  The measurements by the LEP energy group can be cross checked using Z recoil events.
  Preliminary results of the four LEP experiments ALEPH, DELPHI, L3, and OPAL are presented.
  The combination of the results shows no significant deviation of the $E_{\rm beam}$ value obtained using Z
  recoil events from the measurement by the LEP energy group.
}

\section{Determination of the LEP beam energy}
The very precise method to determine the beam energy that was used for the Z run of LEP with an uncertainty of about 1 MeV,
the resonant depolarization method~\cite{Arnaudon:1994zq}, is not available for beam energies above 60 GeV
because the electron and positron beams remain unpolarized beyond this energy.
To determine the energy of beams of a higher energy -- more than 100 GeV in the last year of LEP operation -- another method had to be used.
Nuclear magnetic resonance (NMR) probes were installed in 16 of the LEP dipole magnets.
The beam energy is obtained using the magnetic field $B$ measured by these probes: $E_{\rm beam} \sim \oint B\cdot dl$.
This value has to be calibrated against the energy from the depolarization method at lower energies by assuming
a linear relation between $E_{\rm beam}$ and $B$ to extrapolate to higher energies.
The extrapolation method is cross checked using flux loop measurements, a spectrometer, and spin tune measurements~\cite{lepenergy}.
The uncertainty on the LEP beam energy measurement from these methods is about 20 MeV.
The LEP energy model also describes the time-dependent energy variation at each interaction point.
An independent cross check can be obtained by measuring the kinematics of Z recoil events as described below.

\section{Z recoil events}
\begin{figure}
  \begin{center}
    \includegraphics[scale=.75]{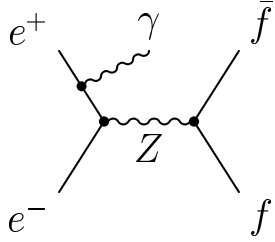}
  \end{center}
  \caption{
    The Feynman diagram for Z recoil events.
    The beam electron or positron emits a hard initial state photon.
    An on-shell Z boson is produced, which decays to a fermion-antifermion-pair.
  }
  \label{fig:eeffg}
\end{figure}

\begin{figure}[!b]

  \vspace{-20pt}
  \begin{center} 
    \includegraphics[scale=.5]{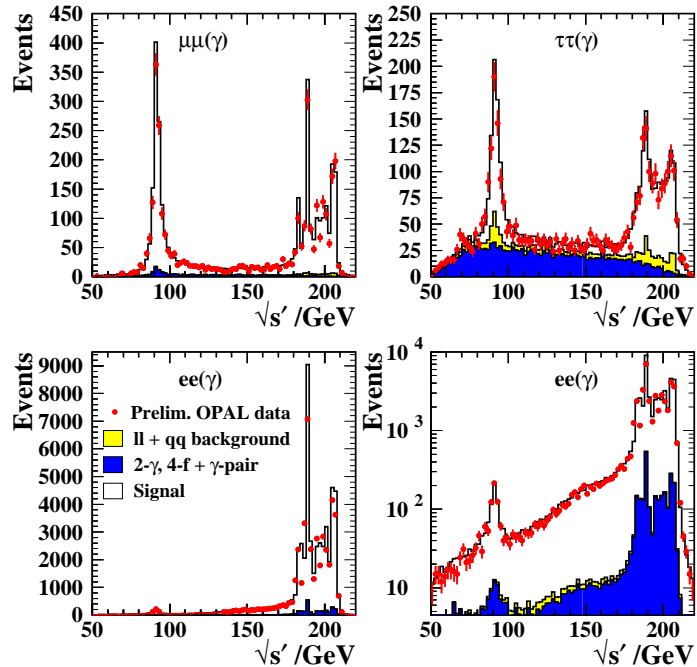}
  \end{center}

  \vspace{-20pt}
  \caption{
    Distributions of the effective centre-of-mass energy, \rtsp, as measured by OPAL, for $\mu^{+}\mu^{-}(\gamma)$ final states (upper left), $\tau^{+}\tau^{-}(\gamma)$ final states (upper right), and $\rm e^{+}e^{-}(\gamma)$ final states (lower plots).
    Due to the large t-channel contribution, the recoil peak for the electron case is not seen as clearly as for the other channels.
    The ``comb'' structure at the end of the spectrum is due to the integral over different energies.
  }
  \label{fig:rtsp}
\end{figure}

Figure~\ref{fig:eeffg} shows the Feynman diagram of the ``radiative return'' events used for the beam energy determination.
Either the incoming electron or positron emits a high energy photon before the collision.
The Z exchange process has a high cross section at energies around the Z mass.
Therefore the energy of the photon is such that an on-shell Z boson is produced in most cases.
In general, the photon escapes along the beam direction, but it is sometimes observed in the calorimeters.

Due to photon emission, the effective centre-of-mass energy, $\sqrt{s^{\prime}}$,
for fermion-antifermion-pair production is reduced from the nominal centre-of-mass energy $\sqrt{s}$.
The reduced centre-of-mass energy can be extracted from event kinematics using
\begin{equation}
  \rtsp{} = \sqrt{s}\cdot\sqrt{\frac{\sin{(\theta_{f})} + \sin{(\theta_{\bar{f}})} - \mid\sin{(\theta_{f}+\theta_{\bar{f}})}\mid}{\sin{(\theta_{f})} + \sin{(\theta_{\bar{f}})} + \mid\sin{(\theta_{f}+\theta_{\bar{f}})}\mid}},
\end{equation}
where ($\theta_{\bar{f}}$) $\theta_{f}$ is the angle between the (anti-)fermion and the photon.
Alternatively, a kinematic fit can be applied to the event, imposing four-momentum-conservation.
This fit significantly enhances the resolution of the spectrum of the effective mass $m_{\rm eff}$.

\begin{figure}
  \begin{center}
    \includegraphics[width=.45\textwidth]{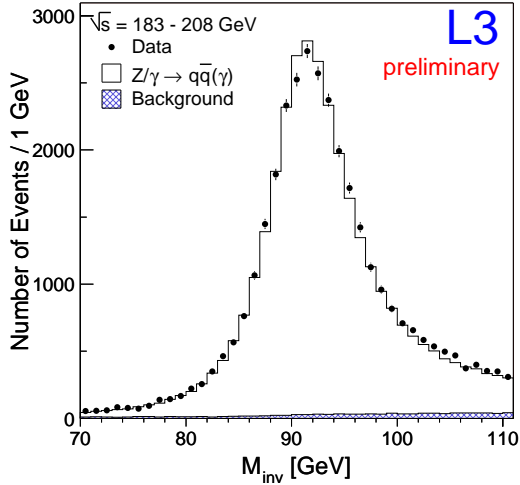}
  \end{center}
  \caption{
    Effective mass of two-jet events, taken with the L3 detector, after kinematic fit, in the recoil peak region.
  }
  \label{fig:l3minv}
\end{figure}

All four LEP experiments use the muon pair final state for this analysis.
This channel has a very clear signature and low background.
OPAL also uses the electron and the tau pair final states.
Figure~\ref{fig:rtsp} shows \rtsp-distributions for lepton final states, as measured by OPAL.
The radiative return peak at effective centre-of-mass energies of the Z boson mass is clearly visible.
In addition, DELPHI, L3, and OPAL use two jet final states, making use of the high statistics of this channel.
Figure~\ref{fig:l3minv} shows the distribution of the effective mass after kinematic fit for the L3 hadronic data.

The analyses presented here~\cite{alephmuon,delphiebeam,l3zmass,opalebeam} are performed
on the 1997 to 2000 data sets with a total integrated luminosity of about 685 $\rm pb^{-1}$ per experiment.

\section{Beam energy extraction}
Different methods are used to extract the beam energy from the \rtsp{} or $m_{\rm eff}$ spectra.
In general, the precise knowledge of the Z boson mass, $m_{\rm Z} = 91.1876 \pm 0.0021$ GeV~\cite{Hagiwara:fs},
allows to measure the beam energy using the position of the radiative return peak.
In the analyses either $E_{\rm beam}$ is assumed to be the nominal value, then a new $m_{\rm Z}$ value is determined, 
and from the deviation of the new Z mass from the precision measurement a shift in the beam energy can be extracted.
Alternatively, the Z boson mass is assumed to be the precision mass, and the beam energy is extracted directly.

For ALEPH muon pair events, a maximum likelihood fit is performed on the distributions of \rtsp{} and on the ``raw'' effective mass
\begin{equation}
  m_{\rm raw}^{2} = 2\cdot p_{1}p_{2} \cdot (1 - \cos\theta_{12}),
\end{equation}
where $p_{i}$ are the measured momenta of the muons and $\theta_{12}$ is the angle between them.

From a likelihood fit to both spectra, Data - Monte Carlo differences $\Delta \rtsp$ and $\Delta m_{\rm raw}$ are extracted.
These values are used to create a $\chi^{2}$ function, which depends on the beam energy.
By minimizing $\chi^{2} = f(E_{\rm beam})$, a possible deviation
\begin{equation}
  \deb = E_{\rm beam}^{\rm meas} - E_{\rm beam}^{\rm LEP}
\end{equation}
from the nominal beam energy is obtained.

\begin{figure}
  \hfill
  \includegraphics[width=.4\textwidth]{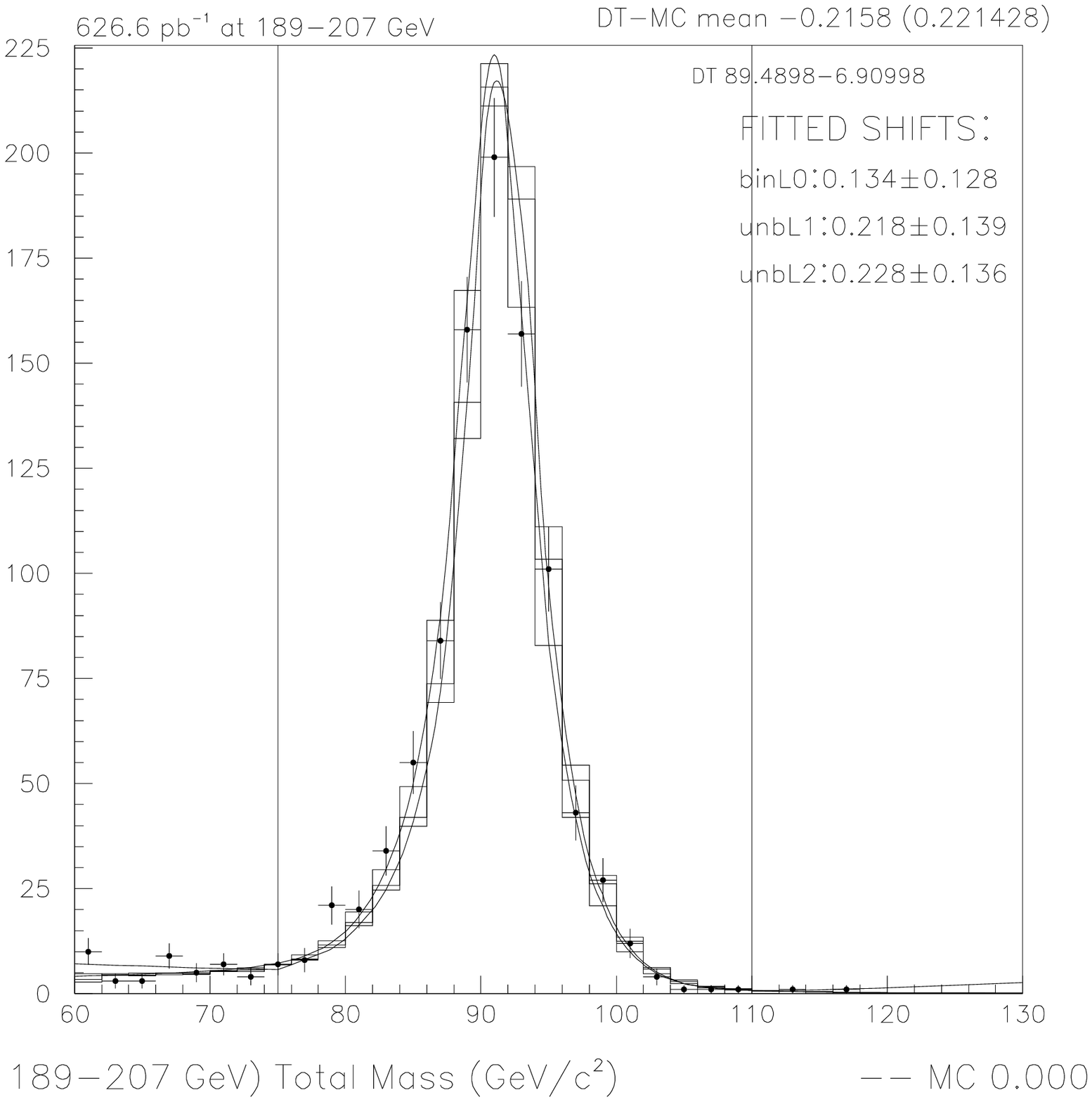}
  \hspace{2cm}
  \includegraphics[width=.4\textwidth]{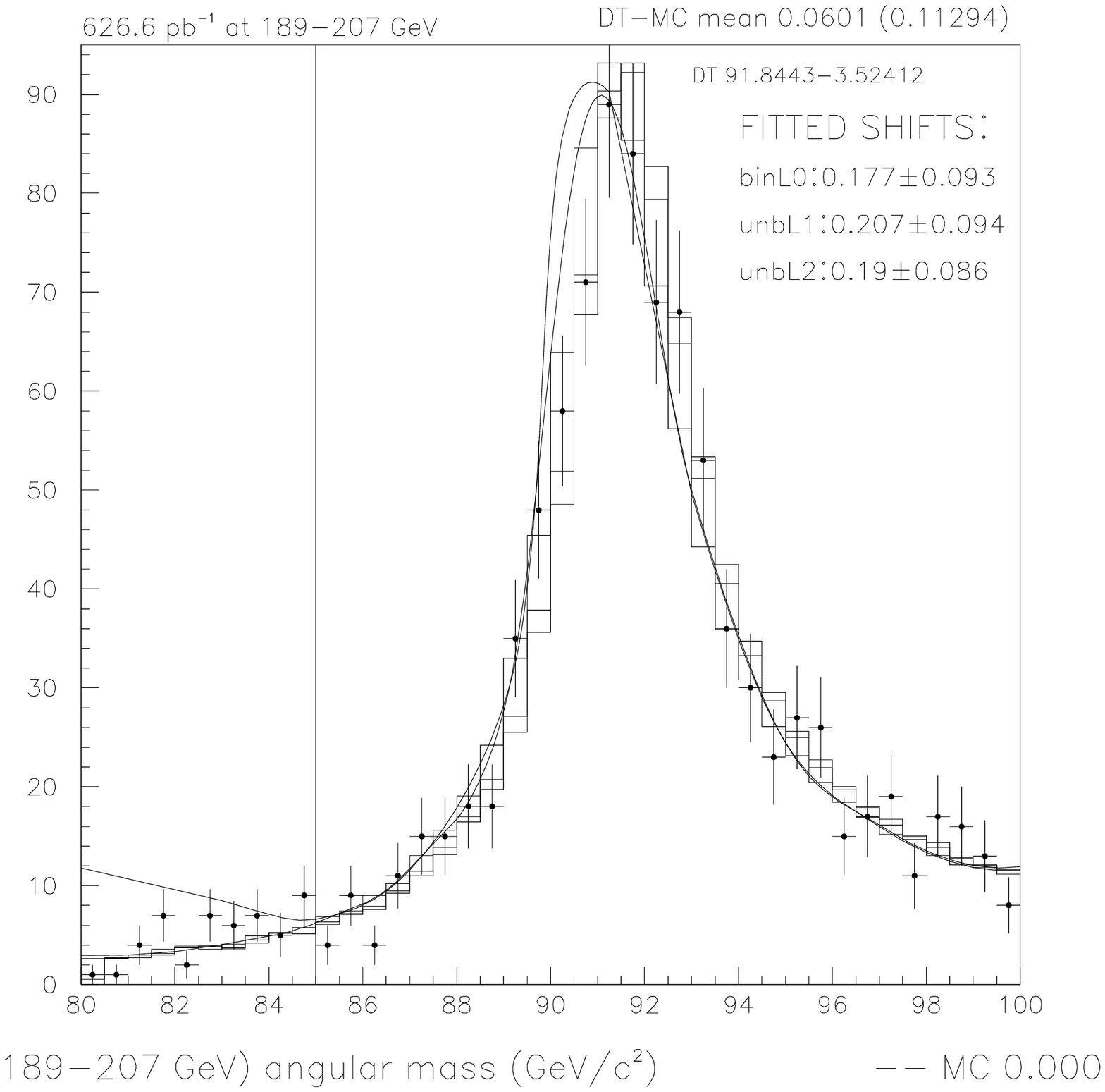}
  \hfill

  \caption{
    The ``raw'' effective mass (left) and the effective centre-of-mass energy (right) of muon pair events, as recorded with the ALEPH detector.
    Dots represent Data, whereas the histograms show Monte Carlo predictions for different Z masses.
    The lines show the functions used in the maximum likelihood fit.
  }
  \label{fig:aleph}
\end{figure}

\begin{figure}
  \includegraphics[width=.4\textwidth]{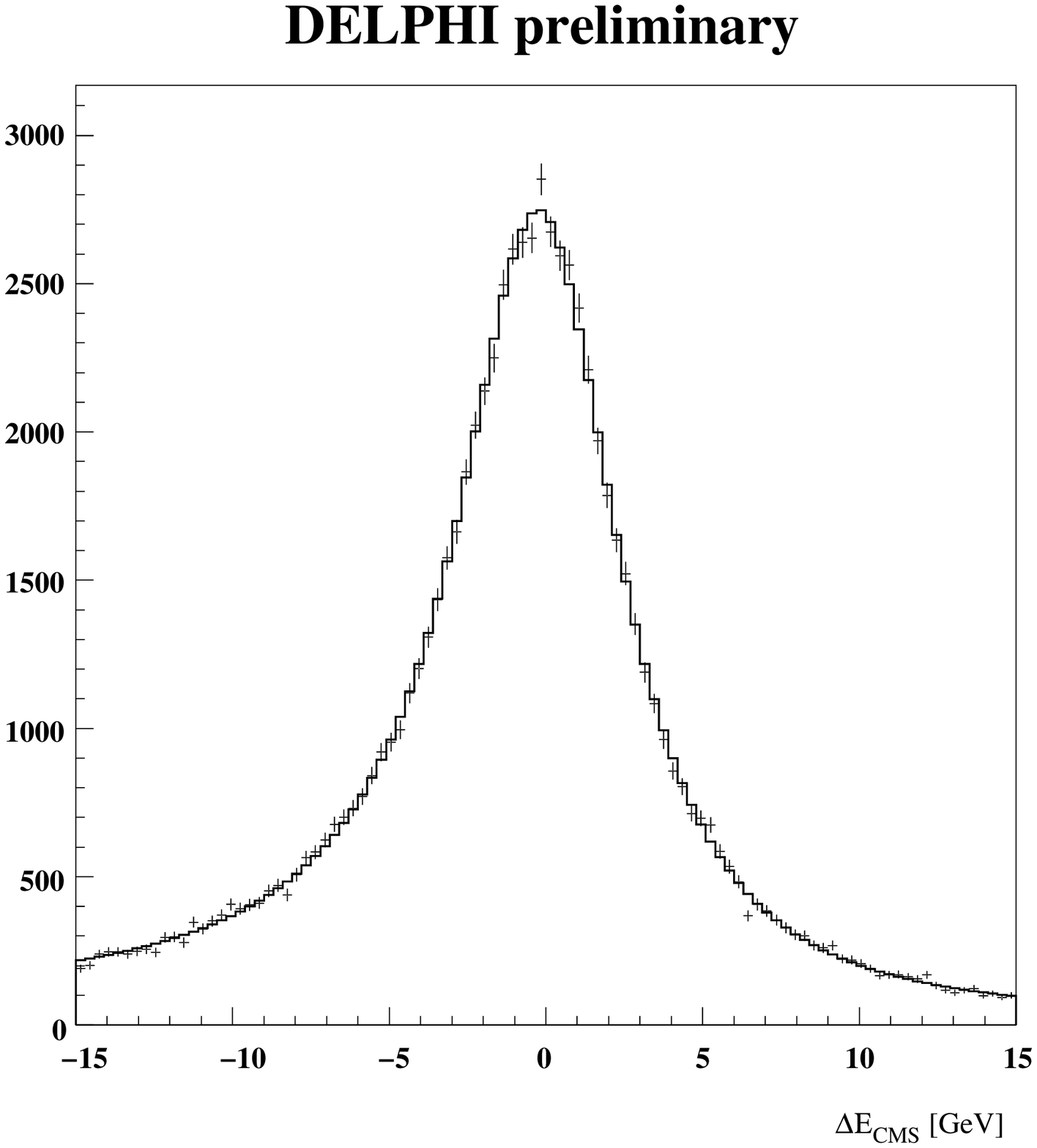}
  \hfill
  \includegraphics[width=.4\textwidth]{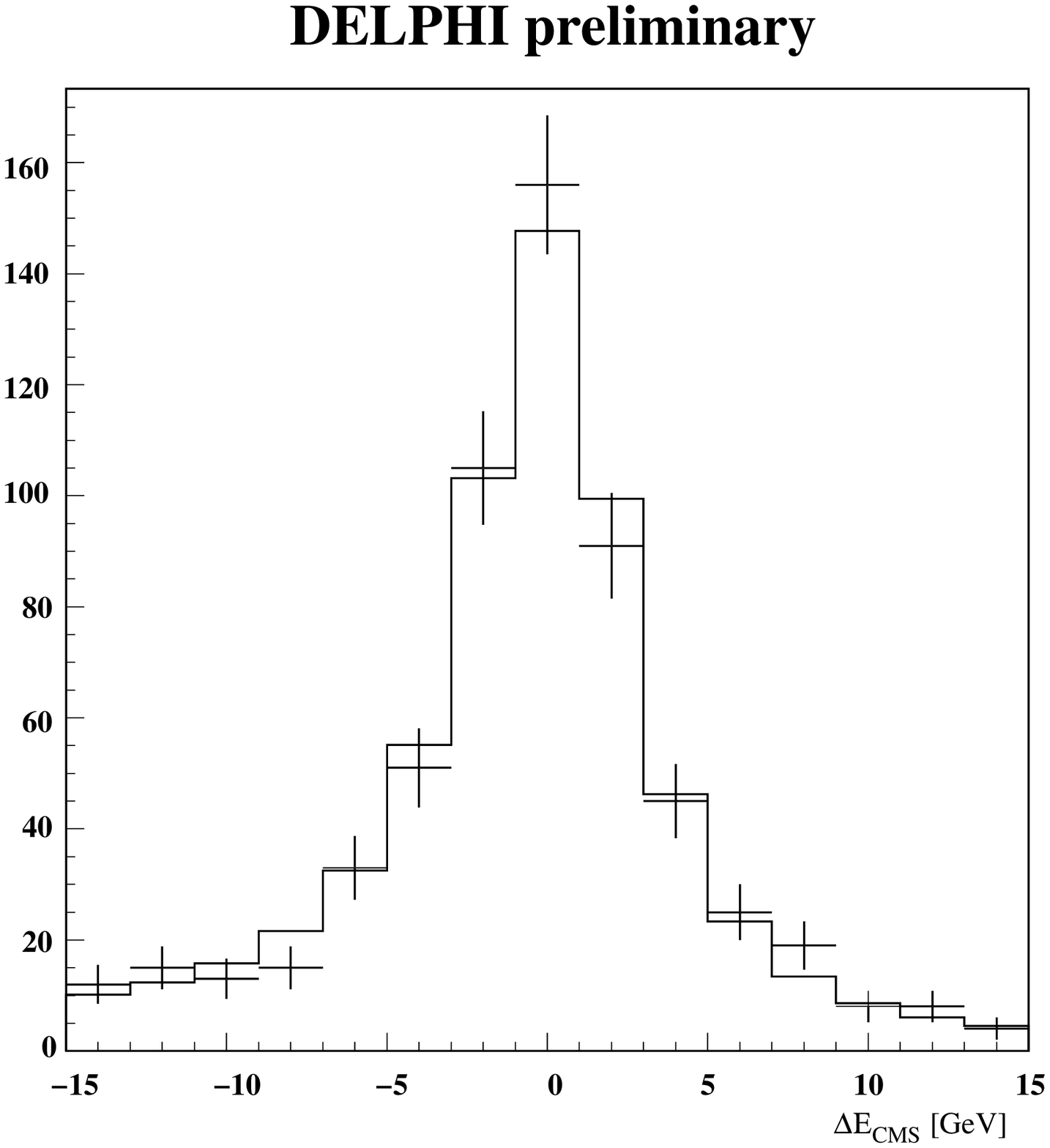}

  \caption{
    $\Delta E_{\rm CM}$ distributions from DELPHI, for Monte Carlo prediction (left) and $\mu$-pair Data (right).
    A Breit-Wigner-like function is fitted to both spectra.
  }
  \label{fig:delphidecm}
\end{figure}

In the second method, used by DELPHI and OPAL for lepton pair events, the Z boson mass is assumed to be the precision mass.
The deviation from the nominal centre-of-mass energy, $\Delta E_{\rm CMS} = 2\cdot\Delta E_{\rm beam}$ is expressed as
\begin{equation}
  \Delta E_{\rm CMS} = \frac{m_{\rm Z}}{x} - \sqrt{s}_{\rm LEP},
\end{equation}
where $x = \frac{\rtsp}{\sqrt{s}}$, a function of the (anti-)fermion-photon angles only.
A possible deviation from the nominal beam energy, \deb, is then extracted from comparing the results 
from a fit of a Breit-Wigner-like function to the $\Delta E_{\rm CMS}$ distributions for Data and Monte Carlo prediction as shown in Figure~\ref{fig:delphidecm}.

In an ansatz used by DELPHI for hadronic final states and by L3 for hadronic and muon pair events, an unbinned maximum likelihood fit, using Monte Carlo event weights
\begin{equation}
  w = \frac{f(\rtsp{}, m_{\rm test})}{f(\rtsp{}, m_{\rm Z})},
\end{equation}
where $f(\rtsp{}, m)$ is a Breit-Wigner-like function and $m_{\rm test}$ is the mass to be tested,
is performed to the \rtsp{} or $m_{\rm eff}$ spectra to extract the Z boson mass, $m_{\rm Z}^{\rm fit}$.
Any deviation from the precision mass is attributed to a deviation of the beam energy from the nominal value using
\begin{equation}
  \deb = - E_{\rm beam}^{\rm LEP}\cdot \frac{m_{\rm Z}^{\rm fit} - m_{\rm Z}}{m_{\rm Z}}.
\end{equation}

In the last method described here, used by OPAL for hadronic events, a Breit-Wigner-like function is fitted
to the Monte Carlo \rtsp{} spectrum to obtain the peak position $m^{*}_{\rm mc}$.
The same function is fitted to the data spectrum, leaving only the normalisation and the peak position, $m^{*}_{\rm Data}$,
as function of $E_{\rm beam}$, as free parameters.
A distribution of the data peak position versus the beam energy is obtained, and the beam energy is extracted by requiring
\begin{equation}
  m^{*}_{\rm Data}(E_{\rm beam}) = m^{*}_{\rm mc}.
\end{equation}

\section{Systematic uncertainties}
The main systematic uncertainties for lepton pair events are the determination of the lepton angles, the ISR modeling 
(studied by using different orders of photon radiation as implemented in the $\mathcal{KK}2f$~\cite{Jadach:2000ir} Monte Carlo generator),
and fit properties.
Further uncertainty sources include background uncertainties, Monte Carlo statistics, and LEP parameters.
The total systematic uncertainty for the beam energy shift observed in leptonic events is 16 MeV.

For hadronic events, the systematic uncertainty is dominated by hadronisation uncertainties.
They are studied by comparing the ARIADNE~\cite{Lonnblad:1992tz}, HERWIG~\cite{Corcella:2002jc}, and PYTHIA~\cite{Sjostrand:2001yu} hadronisation models.
Further major contributions are jet measurements and detector modeling.
For the combined result from hadronic events, the total systematic uncertainty amounts to 43 MeV.

For the combined result, the main contribution to the systematic uncertainty, amounting to 15 MeV, is due to hadronisation.
The total systematic uncertainty is 22 MeV.

\section{Results}
\begin{figure}[!t]
  \begin{center}
    \textsf{LEP preliminary}

    \includegraphics[width=.5\textwidth]{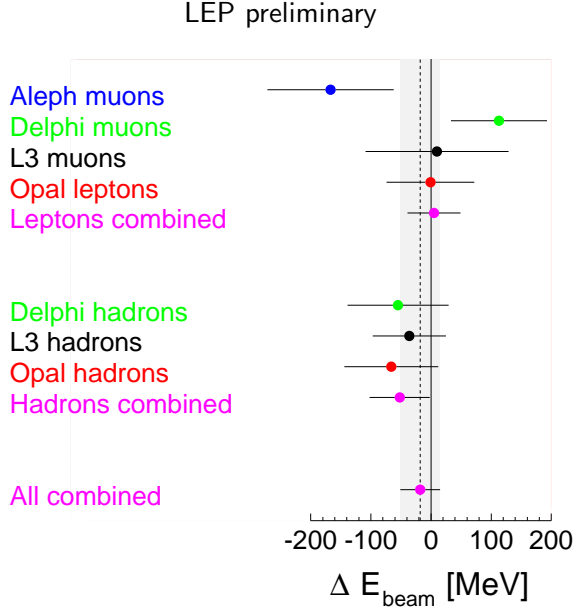}
  \end{center}
  \caption{
    Graphical representation of preliminary $E_{\rm beam}$ measurements performed by the four LEP experiments.
    Shown is the shift from the nominal beam energy.
    The error bars indicate the total error.
  }
  \label{fig:results}
\end{figure}

Table~\ref{tab:res} shows preliminary results of each experiment for leptonic and hadronic final states.
A graphical representation is given in Figure~\ref{fig:results}.
The final result, obtained by averaging the results from lepton pair and hadronic channels of all experiments, taking all correlations into account, yields
\begin{equation}
  \Delta E_{\rm beam} = E_{\rm beam}^{\rm meas} - E_{\rm beam}^{\rm LEP} = (-20 \pm 25 \pm 22) {\rm MeV}.
\end{equation}

This result demonstrates the good agreement between the beam energy measurement from radiative events and the standard LEP energy determination.

\begin{table}[!t]
  \begin{center}
    \begin{tabular}{lcc}
      Experiment & Channel & $\Delta E_{\rm beam}$ [MeV]\\
      \hline
      ALEPH    & $\mu^{+}\mu^{-}\gamma$   & $-167 \pm \phantom{0}91 \pm \phantom{0}48$\\
      DELPHI   & $\mu^{+}\mu^{-}\gamma$   & $+113 \pm \phantom{0}75 \pm \phantom{0}27$\\
      DELPHI   & $\rm q\bar{q}\gamma$     & $-\phantom{0}55 \pm \phantom{0}53 \pm \phantom{0}65$\\
      L3       & $\mu^{+}\mu^{-}\gamma$   & $+\phantom{0}10 \pm 115 \pm \phantom{0}22$\\
      L3       & $\rm q\bar{q}\gamma$     & $-\phantom{0}46 \pm \phantom{0}33 \pm \phantom{0}51$\\
      OPAL     & $\rm e^{+}e^{-}\gamma$   & $+\phantom{0}40 \pm 136 \pm \phantom{0}78$\\
      OPAL     & $\mu^{+}\mu^{-}\gamma$   & $-\phantom{0}51 \pm \phantom{0}84 \pm \phantom{0}22$\\
      OPAL     & $\tau^{+}\tau^{-}\gamma$ & $+301 \pm 199 \pm 148$\\
      OPAL     & $\rm q\bar{q}\gamma$     & $-\phantom{0}66 \pm \phantom{0}34 \pm \phantom{0}70$\\
      \hline
      Combined & $\rm l^{+}l^{-}\gamma$   & $+\phantom{00}5 \pm \phantom{0}41 \pm \phantom{0}16$\\
      Combined & $\rm q\bar{q}\gamma$     & $-\phantom{0}52 \pm \phantom{0}24 \pm \phantom{0}43$\\
      \hline
      Combined & $\rm f\bar{f}\gamma$     & $-\phantom{0}20 \pm \phantom{0}25 \pm \phantom{0}22$
    \end{tabular}
  \end{center}
  \caption{
    $\Delta E_{\rm beam}$ results from each experiment.
    The first error value represents the statistical contribution, the second the systematic contribution.
  }
  \label{tab:res}
\end{table}

\end{document}